\documentclass[twocolumn,superscriptaddress,amsmath,amssymb,prx,floatfix]{revtex4-2}

\usepackage{graphicx}
\usepackage{braket}
\usepackage{dcolumn}
\usepackage{bm}

\usepackage[utf8]{inputenc}
\usepackage[T1]{fontenc}
\usepackage{mathptmx}
\usepackage{etoolbox}

\usepackage[dvipsnames]{xcolor}
\usepackage[normalem]{ulem}
\usepackage{hyperref}
\usepackage{booktabs}

\usepackage{hyperref}
\hypersetup{
    colorlinks=true,
    linkcolor=blue,
    filecolor=magenta,      
    urlcolor=cyan,
    }

\usepackage{amssymb}

\newcommand{\beginsupplement}{%
        \setcounter{table}{0}
        \renewcommand{\thetable}{SM\arabic{table}}%
        \setcounter{figure}{0}
        \renewcommand{\thefigure}{SM\arabic{figure}}%
        \setcounter{equation}{0}
        \renewcommand{\theequation}{SM\arabic{equation}}%
        \setcounter{section}{0}
        \renewcommand{\thesection}{SM\arabic{section}}%
     }

\usepackage{xcolor}
\usepackage[final]{changes}
\definechangesauthor[color=red, name={Paulo Santos}]{PVS}
\definechangesauthor[color=blue, name={Alexander Kuznetsov}]{AK}
\definechangesauthor[color=green, name={Ismael}]{IE}

\setauthormarkup{}

\newcommand{\rAK}[2]{\replaced[id=AK]{#1}{#2}}
\newcommand{\aAK}[1]{\added[id=AK]{#1}}

\newcommand{\rIE}[2]{\replaced[id=AK]{#1}{#2}}

\newcommand{\Pexc}[0]{P_\mathrm{exc}}
\newcommand{\Pth}[0]{P_\mathrm{th}}

\newcommand{\gammaPol}[0]{\gamma}
\newcommand{\gammaBEC}[0]{\gamma}

\makeatletter
\def\@email#1#2{%
 \endgroup
 \patchcmd{\titleblock@produce}
  {\frontmatter@RRAPformat}
  {\frontmatter@RRAPformat{\produce@RRAP{*#1\href{mailto:#2}{#2}}}\frontmatter@RRAPformat}
  {}{}
}%
\makeatother

\newcommand{\citesupp}[1]{\cite{#1}}

\begin{document}





\title{Near-room-temperature zero-dimensional polariton lasers with sub-10 GHz linewidths} 



\author{I.~dePedro-Embid}
\affiliation{Paul-Drude-Institut f\"{u}r Festk\"{o}rperelektronik, Leibniz-Institut im Forschungsverbund Berlin e.V., Hausvogteiplatz 5-7, 10117 Berlin, Germany}
\author{A.~S.~Kuznetsov}
\affiliation{Paul-Drude-Institut f\"{u}r Festk\"{o}rperelektronik, Leibniz-Institut im Forschungsverbund Berlin e.V., Hausvogteiplatz 5-7, 10117 Berlin, Germany}
\author{K.~Biermann}
\affiliation{Paul-Drude-Institut f\"{u}r Festk\"{o}rperelektronik, Leibniz-Institut im Forschungsverbund Berlin e.V., Hausvogteiplatz 5-7, 10117 Berlin, Germany}
\author{A.~Cantarero}
\affiliation{Molecular Science Institute, University of Valencia, PO Box 22085, 46071 Valencia, Spain}
\author{P.~V.~Santos}
\email[]{santos@pdi-berlin.de}
\affiliation{Paul-Drude-Institut f\"{u}r Festk\"{o}rperelektronik, Leibniz-Institut im Forschungsverbund Berlin e.V., Hausvogteiplatz 5-7, 10117 Berlin, Germany}


\date{\today}


\begin{abstract}

Narrow and brilliant spectral lines are essential assets for high-resolution spectroscopy as well as for precision sensing and optomechanics. In semiconductor structures and, in particular, in the well-established (Al,Ga)As material system, strong emission lines with nanosecond coherence times can be provided by the opto-electronic resonances of microcavity exciton-polariton condensates. The temporal coherence of these resonances, however, normally rapidly deteriorates as the temperature increases beyond a few tens of kelvins due to exciton dissociation. Here, we demonstrate that the temperature stability of polariton condensates in (Al,Ga)As can be significantly improved by confinement within $\mu$m-sized intracavity traps. We show that trapped condensates can survive up to $\sim$ 200~K while maintaining a light-matter character with decoherence rates below 10 GHz (i.e., $<40~\mu$eV linewidths). These linewidths are by an order of magnitude smaller than those so far reported for other solid-state systems at these temperatures. Confinement thus provides a pathway towards room-temperature polariton condensation using the well-established (Al,Ga)As material system with prospects for application in  scalable on-chip photonic devices for optical processing, sensing, and computing. 

\end{abstract}

\maketitle 

\begin{figure*}[tbhp]
    \includegraphics*[width=1\textwidth]{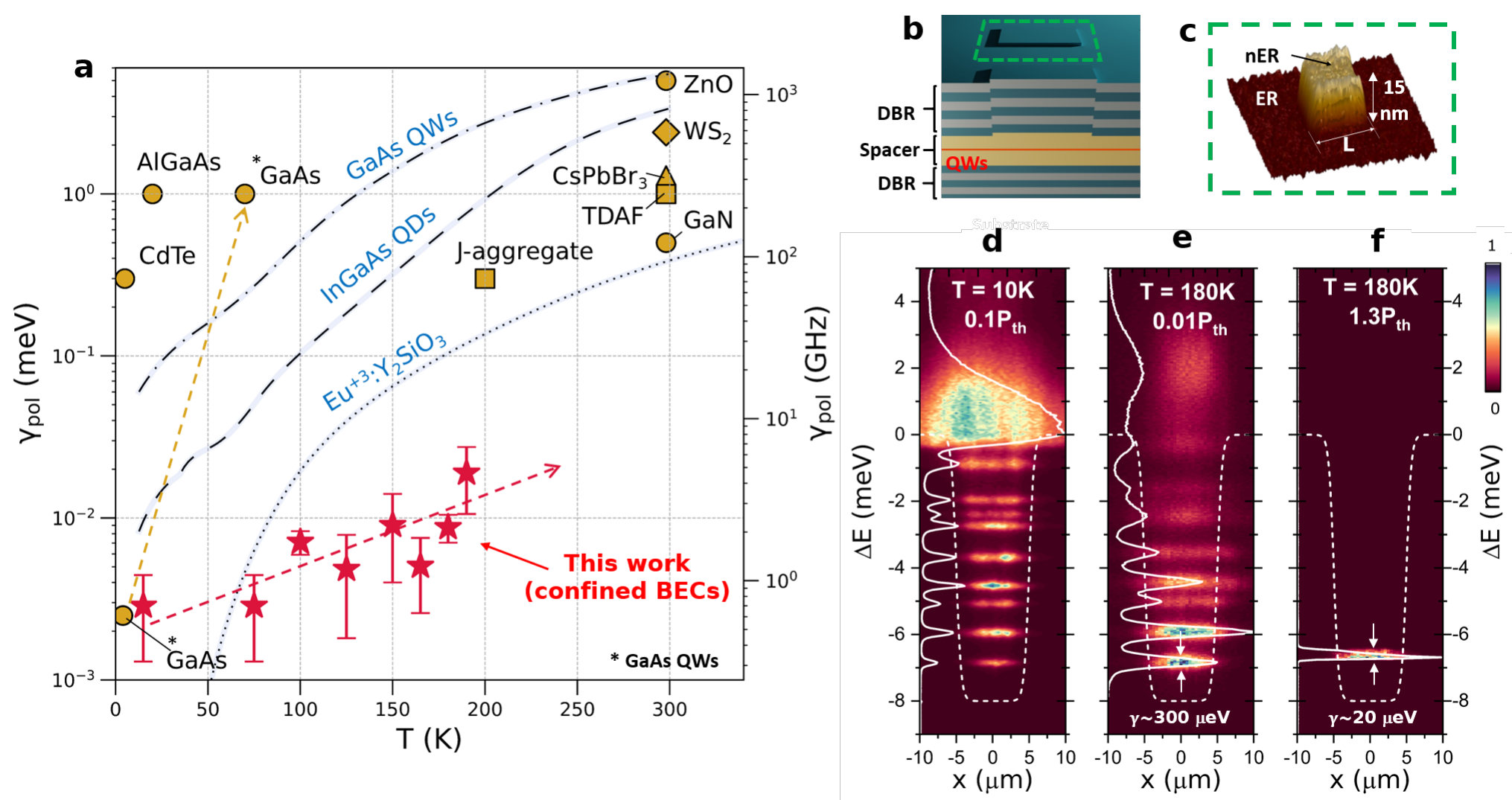}
    \caption{    
           {\bf Intracavity traps for polaritons}:      
             a. Reported linewidths  $\gamma$ for  GaAs quantum wells (QWs, dot-dashed line), InGaAs quantum dots (QD, dashed line), yttrium silicate color centers (Eu$^{+3}$:Y$_2$SiO$_5$, dotted line) as well as for polariton Bose-Einstein condensates (BECs) on different material systems 
            ($\textcolor{orange}{\bullet}$: II-VI and III-V semiconductors,
             $\textcolor{orange}{\blacksquare}$: organic semiconductors, 
             $\textcolor{orange}{\blacklozenge}$: transition-metal dichalcogenides (TMDC), 
             $\textcolor{orange}{\blacktriangle}$: halide perovskites). 
             The   
             $\textcolor{red}{\bigstar}$'s are the corresponding linewidths for the GaAs-based confined BECs  introduced in this work (further details in the supplement, Sec.~\ref{sec:SM_Pol_FWHM_Rev}).
             b. Cross-section of a structured (Al, Ga)As polariton microcavity (MC) with intracavity traps.
             The MC consists of two distributed Bragg reflectors (DBRs) enclosing a spacer with GaAs quantum wells (QWs). 
             Lateral variations of the spacer thickness by etching result in lower polariton energies in the non-etched regions (nER) than in the surrounding etched ones (ER), thus creating an intracavity polariton trap. 
             c. Atomic force micrograph recorded on the surface of a structured MC with a $4\times 4~\mu$m$^2$  intracavity trap.  
             d-e. Spatially resolved photoluminescence (PL) map of a $5\times 5~\mu$m$^2$  intracavity trap recorded at 10~K  and 180~K, respectively, for optical excitations, $\Pexc$, below the BEC threshold, $\Pth$. The PL energies ($\Delta E$) are relative to the trap barrier. 
             f.  Corresponding map for the trap in $e$ in the BEC  regime. The color scale is normalized PL and  the energy scale referenced to the trap barriers.  The superimposed white lines yield the spatially integrated spectra, the dashed ones sketch the lateral trap confinement potential. 
            }
    \label{FWHM_Rev}
    \end{figure*}

Spectrally narrow opto-electronic resonances form the basis for compact integration of a wide range of functionalities including advanced sensing, laser frequency stabilization as well as optical signal processing and storage. The optical nature of these resonances enables fast \rAK{readout }{access} while the electronic component provides the required sensitivity to external fields demanded for sensing and control. Figure~\ref{FWHM_Rev}a compiles the linewidths ($\gamma$, proportional to the decoherence rates) reported for \rAK{well-studied }{narrow} spectroscopic resonances in the solid-state of optoelectronic nature (i.e., excluding purely photonic lasing). Among the narrowest, one finds rare-earth related color centers, some with homogeneous linewidths   $\gamma < 100$~Hz at cryogenic temperatures (e.g. europium centers in yttrium silicate,  Eu$^{+3}$:Y$_2$SiO$_5$, dotted line in Fig.~\ref{FWHM_Rev}a~\cite{Koenz_PRB68_85109_03}). The record narrow linewidths of these centres are at least one order of magnitude below  those reported for semiconductor quantum wells (QWs, cf.~dot-dashed line in Fig.~\ref{FWHM_Rev}a) and quantum dots~\cite{Bayer_PRB65_041308_02} (QDs, dashed line). Other examples of spectrally narrow color center resonances are provided by phosphorus impurity in Si ($<5$~MHz~\cite{Yang_APL95_09}),  nitrogen vacancy centers in diamond (13~MHz~\cite{Tamarat_PRL97_83002_06}) and silicon-vacancy centers in SiC (70~MHz~\cite{Nagy_NC10_1954_19}). 

An alternative approach towards narrow solid-state resonances explores microcavity exciton-polaritons (polaritons hereafter) -- hybrid light-matter quasi-particles arising from the strong coupling between quantum well (QW) excitons and photons confined in a semiconductor microcavity (MC)~\cite{Weisbuch_PRL69_3314_92}. Polaritons combine the long temporal and spatial coherences inherited from their photonic character with the strong nonlinearities and efficient coupling to solid-state excitations mediated by their excitonic component, thus forming an exceptional bridge between condensed matter and photonic systems. Furthermore, the  bosonic nature of the polaritons together with their small effective mass ($\sim10^{-4}-10^{-5}$ of the free electron mass) enable Bose-Einstein condensation (BEC)~\cite{Kasprzak_N443_409_06} and polariton lasing~\cite{Deng_PNAS100_15318_03} even at room temperature (RT)~\cite{Christopoulos_PRL98_126405_07}. 
The orange symbols of Fig.~\ref{FWHM_Rev}a summarize reported linewidths for polariton lasers in different material systems. At low temperatures, the linewidths of (Al,Ga)As polariton lasers, the material platform presently offering the highest degree of growth control and structural perfection~\cite{Biermann_JCG557_125993_21}, reach values  down to 54~MHz~\cite{Fabricante_O11_838_24}. They are, thus, among the narrowest aforementioned \aAK{optical} resonances.
For other material platforms, the emission linewidths typically exceed 73~GHz (300~$\mu$eV). The remarkable coherence and non-linearities  make polaritons ideal candidates for a number of novel optoelectronic device functionalitiess~\cite{Sanvitto_NM15_1061_16}.

A major challenge towards applications of all \aAK{spectrally} sharp solid-state  resonances of  Fig.~\ref{FWHM_Rev}a is their limitation to low  temperatures, i.e., typically  T$<10$~K. As illustrated by Fig.~\ref{FWHM_Rev}a,  they significantly broaden above a few 10's of K mainly due to interactions with phonons. For the  (Al,Ga)As polariton platform, condensation has been reported up to approximately 77~K\cite{Tempel_NJP14_83014_12}. Recently, long-range spatial coherence has been reported for this system at room temperatures albeit with relatively large ($>48$~GHz$=200~\mu$eV) spectral linewidths~\cite{Alnatah_AP12_48_24}.

In this paper, we demonstrate bright and highly  coherent polariton lasing in (Al,Ga)As structures reaching spectral  linewidths $<7~$GHz  at temperatures up to 200~K. These linewidths, which are indicated by the  red stars in Fig.~\ref{FWHM_Rev}a, have been measured in polariton BECs confined within $\mu$m-sized intracavity traps fabricated  in a structured  (Al,As)Ga MC~\cite{Schneider_RPP_16503_17, PVS312} (cf. Figs.~\ref{FWHM_Rev}b and \ref{FWHM_Rev}c, see Methods). 
Low temperature ($T\leq 10$~K) photoluminescence (PL) maps recorded under low optical excitation  [cf.~Fig.~\ref{FWHM_Rev}d] display well-defined polariton states confined within the traps, as expected for a quantum particle in a box. These states have linewidths  $\gammaPol \gtrapprox 60$~GHz ($250~\mu$eV) and are surrounded by extended barrier states with much broader emission lines.
Remarkably, the confined states remain stable with essentially the same $\gammaPol$ for the  ground state (GS) up to 200~K, thus demonstrating that confinement significantly enhances temporal coherence [cf.~Fig.~\ref{FWHM_Rev}e]. More importantly, by increasing the particle density, the confined polaritons transition to a bright and highly coherent BEC-like state with  $\gamma<7$~GHz (cf. Fig.~\ref{FWHM_Rev}f).
These temperatures for highly coherent polariton lasing are over twice as large as the highest ones (of approx. 70~K~\cite{Tempel_NJP14_83014_12}) so far reported for 
GaAs-based polariton BECs. The  BEC linewidths at 200~K also subseed those of all other opto-electronic resonances in Fig.~\ref{FWHM_Rev}a by over an order of magnitude. We provide experimental evidence supporting the hybrid light-matter  (rather than purely photonic) nature of the laser-like emission at high temperatures.
Lateral confinement thus significantly \rAK{increases }{broadens} the  temperature stability range over which the exceptional properties of GaAs-based polariton condensates are maintained, thus paving the way for (Al, Ga)As-based polariton lasers operating at temperatures approaching room temperature~\cite{Sanvitto_NM15_1061_16}.

 \begin{figure*}[tbhp]
    \includegraphics*[width=\textwidth]{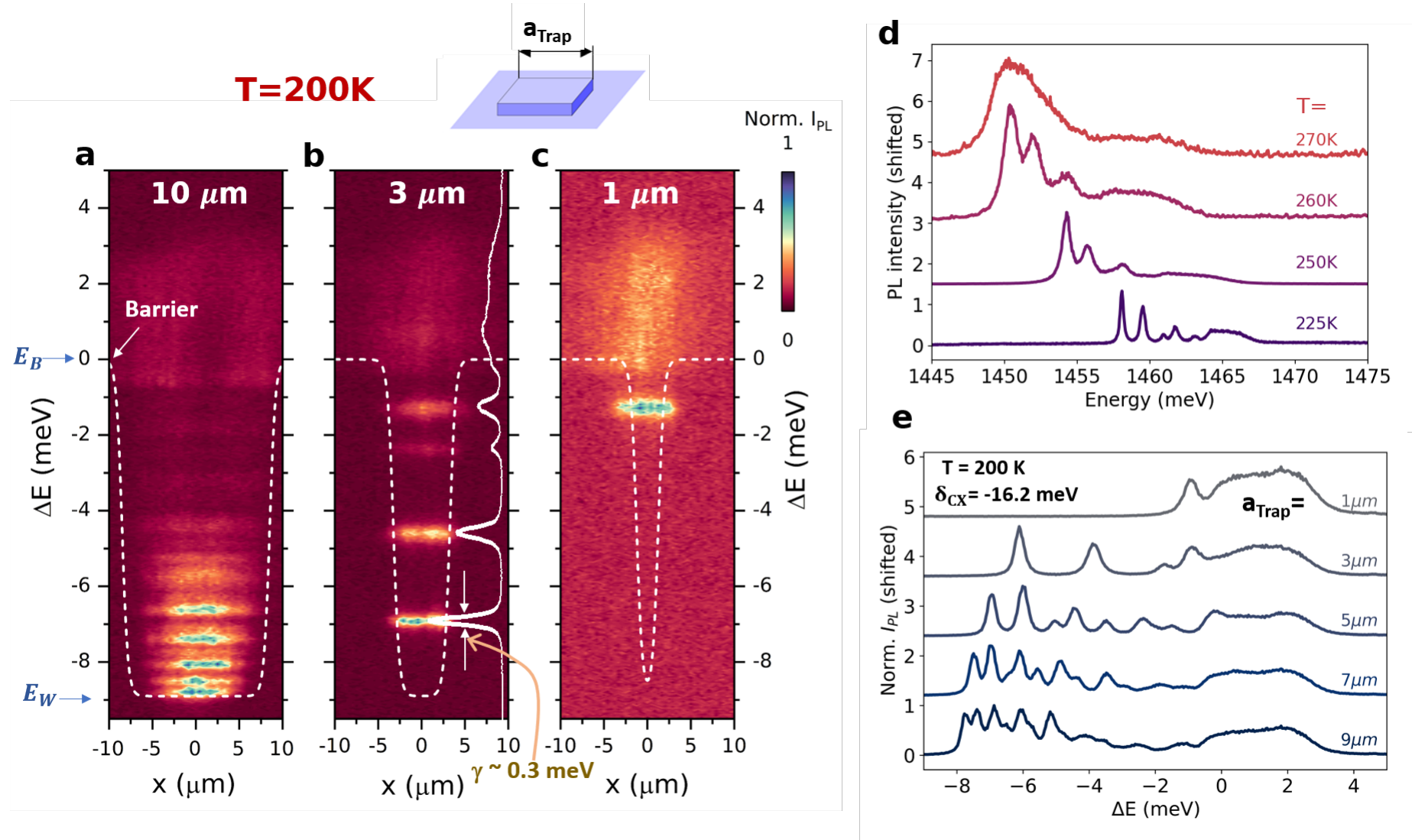}
    \caption
        { {\bf Confined polariton states in the sub-threshold regime.} 
        a.-c.  Spatially resolved PL maps recorded at 200 K on square intracavity traps with sizes $a_\mathrm{Trap}=10$, $3$ and $1~\mu$m, respectively. The color maps are normalized to the maximum PL and the energies  referenced to the barrier energy, $E_B$. The white dashed lines sketch the  confinement potential with bottom energy $E_W$. The superimposed white line in $b$ displays the spatially integrated PL.
        d. Temperature dependence of the PL of a  $3\times 3~\mu$m$^2$ trap. 
        e. PL spectra recorded at 200~K for traps with different sizes and  GS detuning  $\delta_\mathrm{CX}=-16.2$~meV.  
        }
    \label{fig:HT_Conf_Pol}
    \end{figure*}

\section{Results}

\subsection{Confined polaritons states at high temperatures}

The impact of confinement on polariton is illustrated by the PL maps recorded on square traps with different sizes in Figs.~\ref{fig:HT_Conf_Pol}a-c (details of trap fabrication are found in  Methods and in the Supplementary Material (SM), Sec.~\ref{subsec:SM_MC_Structure}). The maps were recorded at 200~K using a low optical excitation power, $\Pexc$  (i.e., well below the BEC excitation threshold, $\Pth$). Similarly to the $5\times 5~\mu$m$^2$ trap of Fig.~\ref{FWHM_Rev}e, the confined polaritons exhibit  discrete states within the trapping potential defined by the lower polariton  energies in extended non-etched and etched areas of the sample  [cf.~Fig.~\ref{FWHM_Rev}b] respectively. As expected, the energy spacing between confined states reduces with increasing trap sizes.
For all trap sizes, the spectral widths $\gammaPol$ of the confined polariton lines are substantially narrower than that for unconfined polaritons in the extended regions of the barrier [see also Fig.~\ref{fig:HT_Conf_Pol}e], thus showing that confinement substantially reduces the polariton decoherence rate.
For all trap sizes, the spectral widths $\gammaPol$ of the confined polariton lines are substantially narrower than that for unconfined polaritons in the extended regions of the barrier [see also Fig.~\ref{fig:HT_Conf_Pol}e], thus showing that confinement substantially reduces the polariton decoherence rate.

The PL temperature dependence  is illustrated for  a $3\times3~\mu$m$^2$ trap in Fig.~\ref{fig:HT_Conf_Pol}d. 
The red-shift of the resonances with temperature is mainly dictated by the strong temperature dependence of the bare excitonic resonances, which far exceed the photonic ones. Polaritons in this trap have a strong photonic character with a cavity-to-exciton detuning $\delta_\mathrm{CX}=-16.2$~meV at the lowest temperature (225~K). The excitonic content (and, thus, $\delta_\mathrm{CX}$) increases as the bare excitonic level red-shifts with increasing temperature [see Sec.~\ref{subsec:SM_MC_Tdep}], which also reduces the trap confinement barrier (the barrier states are also polaritonic).  Both effects lead to a broadening of the confined levels until they become completely washed out  above 270~K. We will show below that the broadening can be partially reverted by increasing $\Pexc$.


\begin{figure*}[tbhp]
    \includegraphics[width=1\textwidth]{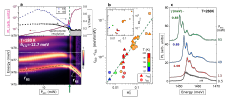}
    \caption
    {
        {\bf Impact of the excitation density on polariton coherence.} 
	    a.  PL  {\it vs.}  optical excitation power ($\Pexc$) at 180~K for a $5\times 5~\mu$m$^2$ intracavity trap with a nominal detuning $\delta_\mathrm{CX}=-18$~meV. The dashed green line  marks the \rIE{lasing}{BEC} threshold $\Pth = 40$~mW. 
        The PL is color-encoded in a normalized log scale. 
        The upper panel shows the dependence on $\Pexc$ of the integrated GS PL (red line) and of the GS (black dots) and first ES (blue triangles) linewidths detected by a conventional spectrometer. 
        b. Dependence on the  exciton Hopfield coefficient, $H^2_X$,  of the rates $\delta E_{GS}/\delta \Pexc$ for the GS as measured  (cf.~dashed black lines in panel a)  for  low ($r_\mathrm{BS}$) and high ($r_\mathrm{BS}$) $\Pexc$. The green dashed line displays an energy shift rate proportional to $H^2_X$. The inset shows the corresponding dependence for the maximum blue-shift (BS$_{max}$).
	    c. Sub-threshold PL spectra of a $4\times 4~\mu$m$^2$ trap recorded at 260~K under different  $\Pexc$. 
        }
    \label{fig:FluenceDep}
    \end{figure*}

\begin{figure*}[tbhp]
    \includegraphics[width=\textwidth]{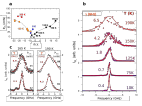}
    \caption
        { {\bf High-temperature polariton condensation. }
		a. BEC excitation threshold $\Pth$ {\it vs.}  detuning $\delta_\mathrm{CX}$ for confined BECs at different temperatures (encoded in the color of the symbols).       
        b. High-resolution PL spectra of  BECs in a $4\times 4~\mu$m$^2$ trap recorded at different temperatures. The energy scale is referenced to the main emission line. The red lines are fits yielding the listed linewidths.
		c. Dependence on $\Pexc$  of the BEC spectra of a $4\times 4~\mu$m$^2$ trap  at 165~K and 190~K.
        }
    \label{fig:BECCoherence}
\end{figure*}

\subsection{Many-body effects on polariton coherence}

The confined polaritons  undergo condensation at high temperatures (i.e., above 100~K) under high particle densities. The color map of Fig.~\ref{fig:FluenceDep}a shows the PL dependence on  $\Pexc$ recorded at 190~K on a $5\times 5~\mu$m$^2$ square trap. The upper panel displays  the integrated  PL intensity for the  GS (red line) as well as the linewidths of the GS (black dots)  and first excited state (ES, blue triangles). The PL intensity increases with $\Pexc$ until one reaches the BEC threshold at $\Pth = 40$~mW. Here, 
the emission from the GS abruptly increases by two orders of magnitude (cf.~red line in the upper panel) and the GS linewidth reduces below the resolution limit.

The polariton dynamics in the sub-threshold regime of Fig.~\ref{fig:FluenceDep}a (i.e., for  $\Pexc<\Pth$)  yields important information about the decoherence mechanisms. The blue-shift with increasing $\Pexc$, which is also observed  at lower temperatures (i.e., for $T<100$~K, cf.~Sec.~\ref{sec:SM_Pexc}),  is attributed to polariton-polariton interactions at increased particle densities. 
The blue-shift at high temperatures, however, reaches a maximum well below $\Pth$ and is then followed by a red-shift, which extends into the BEC regime. 
Additional experiments (cf. Sec.~\ref{sec:SMTEffects}) prove that observed red-shifts at high excitations are of opto-electronic nature rather than due to optical heating.

In the traps studied here, the exciton content quantified by the Hopfield coefficient $H^2_X$ is small and increases for the higher lying confined levels. The stronger blue and red shifts for the ES in the upper panel of Fig.~\ref{fig:FluenceDep}a indicate that both shifts are associated with  $H^2_X$. To further corroborate  the electronic nature of the $\Pexc$-induced shifts, we define rates for the red ($r_\mathrm{RS}$) and  blue ($r_\mathrm{BS}$) energy shifts as derivatives $\delta E_{GS}/\delta \Pexc$ calculated in the regimes of low and high $\Pexc$, respectively (cf.~Fig.~\ref{fig:FluenceDep}a). Figure~\ref{fig:FluenceDep}b shows that both  $r_\mathrm{RS}$ (circles) and  $r_\mathrm{BS}$ (red diamonds) are proportional to  $H^2_X$, thus corroborating their electronic nature. The inset shows that the maximum measured blue-shift (BS$_{max}$) also increases with $H^2_X$.

Finally,  another remarkable feature indicated by the symbols  in Fig.~\ref{fig:FluenceDep}a (upper panel) is the substantial linewidth narrowing with increasing  $\Pexc$, which starts already well below condensation. The line narrowing is relatively small below 180~K (cf. Fig.~\ref{fig:FluenceDep}a): it becomes, however,  very pronounced at higher temperatures. As an example, Fig.~\ref{fig:FluenceDep}c compare subthreshold PL spectra of a $4\times 4~\mu$m$^2$ trap recorded under different $\Pexc$. 
At the lowest power $\Pexc$ (0.5~mW), the confined states cannot be distinguished since their  linewidths is comparable to the depth of the confinement potential.
 The linewidths, however, narrow within increasing $\Pexc$ to values below 1~meV, thus enabling the clear identification of the  confined states. 

\subsection{Polariton lasing threshold at high temperatures}
We now turn our attention to the BEC regime at high excitation densities (i. e., for $\Pexc\ge P_\mathrm{th}$, cf. Fig.~\ref{fig:FluenceDep}a). 
While the emission intensity increases and the linewidth substantially reduces across $P_\mathrm{th}$, the energies of the GS and ESs continuously evolve  across the transition while remaining close to the corresponding ones  at very low optical excitation  (i.e., blue-shifted by less than 0.5 meV). These features are qualitatively similar to observations at lower temperatures (i.e., for $T<100$~K, cf. Sec.~\ref{sec:SM_Pexc}). As will be further justified below, this behavior at high temperatures is attributed to the onset of stimulated scattering to the GS resulting in the formation of a polariton BEC. 

A further hint for the polariton BEC nature of the transition comes from the dependence of $\Pth$ on temperature and on  detuning, $\delta_\mathrm{CX}$, displayed by symbols in Fig.~\ref{fig:BECCoherence}a. Here, the  temperature is encoded in the color of the  symbols. $\Pth$ is mainly determined by $\delta_\mathrm{CX}$ (rather than by temperature) following a single trend with a minimum for  $\delta_\mathrm{CX}\approx -5$~meV close to the exciton-polariton Rabi splitting.

\subsection{Coherence of high-temperature polariton lasers}

The BEC temporal coherence  was accessed via high-resolution (approx. 0.3 GHz) spectroscopy using an etalon (cf. Methods). The experiments proved to be challenging due to required temperature stability as well as to the eventual onset of fluctuations in emission intensity and energy close to $P_\mathrm{th}$ (see Sec.~\ref{sec:SMInstabilities} for details).
Figure~\ref{fig:BECCoherence}b compares the temperature dependence of the GS PL of a $4\times 4~\mu$m$^2$ recorded for $\Pexc$ slightly abover $\Pth$ (symbols). The superimposed red lines are fits to a Lorentzian line shape yielding the listed linewidths $\gammaBEC$. The  PL lines below 100~K have a well-defined Lorentzian shape with linewidths below one GHz, which broaden  with temperature to reach 6.5 GHz at 190~K. The line shapes above 150~K are, in some cases, partially distorted by the previously mentioned instabilities.  Linewidths extracted from similar spectra recorded at different temperatures are displayed as red stars in the overview plot of Fig.~\ref{FWHM_Rev}a (for further details, see Sec.~\ref{sec:SM_Pol_FWHM_Rev}). 

Finally, Fig.~\ref{fig:BECCoherence}c illustrates the impact of  $\Pexc$  on the BEC coherence at  at 165~K and 190~K.  Here, the lower spectra were recorded just above threshold, where the coherence is the longest. For higher excitation, one observes a broadening of the BEC line signalizing a reduction in temporal coherence. 
 
\section{Discussions}
\label{Discussions_and_conclusions}

The main result of the present investigations is the demonstration of narrow ($<10$~GHz) spectral lines from GaAs-based confined polarition BECs up to $\sim200$~K.  These long coherences are found in a regime where the  thermal energies far exceed  the estimated exciton binding energy $E_B= 9~\text{meV} \sim k_B 100$~K of the 15 nm-wide QWs \cite{Andreani_PRB42_8928_90}. The narrow lines enable the observation of well-defined confined states in the sub-threshold regime. In the BEC regime, the polariton linewidths reduce below 10~GHz, i.e., to values
an order of magnitude below those so far reported for other solid-state excitonic systems [cf.~Fig.~\ref{FWHM_Rev}a].  In the following, we first analyse the mechanisms for polariton coherence in the low density regime and then proceed to the BEC regime.


A remarkable feature of the confined polaritons at high temperatures is the strong dependence of the linewidth on excitation density $\Pexc$. The line narrowing starts for $\Pexc$ well below the condensation threshold (cf. Fig.~\ref{fig:FluenceDep}a, upper panel), a trend that continues beyond the BEC transition.  The coherence enhancement with $\Pexc$ comes together with an emission red-shift which, as  the line narrowing, depends on the polariton excitonic content (cf. Fig.~\ref{fig:FluenceDep}b).  Interestingly, Fig.~\ref{fig:FluenceDep}b shows that the excitation-induced  coherence enhancement enables the formation of well-defined confined states at temperatures well above 200 K, where the decoherence at low excitation far exceeds the energy splitting between confined states. 

The coherence enhancement with increasing $\Pexc$  could potentially be explained by a  progressive transition to the weak coupling regime, where excitons partially decouple from photons to enter a 
photonic lasing regime similar to  a vertical cavity emitting laser. Alternatively, the coherence enhancement could arise from  photon condensation, a mechanism recently invoked  for laser-like emission from MCs at room temperature\cite{Pieczarka_NP18_1096_24,Schofield_NP18_1089_24}. 
A  transition to weak-coupling with purely photonic emission is, however, incompatible with several experimental observations, as detailed below. 

First, the emission energies at high-temperatures exhibit a continuous dependence $\Pexc$  accross the  lasing threshold and beyond. The energy shifts with $\Pexc$ correlate very well with the detuning  $\delta_{CX}$, a property  intimately related to strong light-matter coupling. 
Furthermore, Sec.~\ref{sec:Continuity of the polariton energy accross the lasing threshold} shows that  changes in the effective mass due a transition to photonic lasing at the threshold would be accompanied by large energy shifts of the levels, which are  not detected in Fig.~\ref{fig:FluenceDep}. 
Also, a potential transition to the photon lasing regime beyond $\Pth$ has not been observed within the available excitation densities. 

The  lasing dynamics above threshold provides additional support for the polariton nature of the lasing transition. The instabilities observed around the lasing transition (cf.~Sec.~\ref{sec:SMInstabilities}) point to a strong self-interactions as well as with phonons. More importantly, while the coherence of a photonic laser increases proportionally to excitation fluence, we observe exactly the opposite behavior, namely an increase of the linewidth with increasing particle densities above the threshold (cf. Fig.~\ref{fig:BECCoherence}). These findings agree with previous reports for polariton lasing at low temperatures: they are attributed to the increased role of self-interactions at high particle densities~\cite{Porras_PRB67_161310_03}.


Rather than to weak coupling, the long temporal coherences are most likely due  to the enhanced stability of the polariton BECs under lateral confinement. Brodbeck {\it et. al.}~\cite{Brodbeck_PRL119_27401_17} have recently demonstrated that the coupling to the light field in a planar MC effectively  enhances electron-hole attraction via photon-mediated interactions. This regime of   {\it very strong coupling}  sets in already for ratios $\Omega_R/E_B > 0.5$ between the Rabi coupling and the exciton binding energy, which are comparable to the ones for our samples. The lateral confinement within intracavity traps should help towards reaching this regime\cite{Zhang_PRB87_115303_13}. The nature of the involved electronic (or excitonic) complexes at high densities and temperatures may, however,  be different from the Wannier exciton picture at cryogenic temperatures. 

The highest $\Pexc$ in Fig.~\ref{fig:FluenceDep}b ($\sim100$~mW, cf. Fig.~\ref{fig:FluenceDep}b) creates photo-excited densities of electron-hole pairs estimated to be in the range $n_{eh}=10^{12}-10^{13}$~cm$^{-2}$.
The red-shifts observed at high densities are most likely due to carrier-induced band-gap screening~\cite{Fehrenbach_PRL49_1281_82}. 
Furthermore, excitons, as well as loosely bounded electron-hole species, can  at high densities effectively screen the piezoelectric field induced by longitudinal optical (LO) and acoustic (LA) phonons, which provides the dominant phonon scattering mechanism. While a quantitative analysis of this dynamic screening is missing  for confined polaritons, we provide here a rough estimation assuming the presence at high excitations of a free electron density $n_e$ comparable to $n_{eh}$ (created, e.g., by pair dissociation at high temperatures). The screening efficiency depends on the carrier plasma energy, $\omega_p$: phonons with energies below $\omega_p$ are efficiently screened while those above are not. At high $n_e$, $\omega_p$  for the 2D QW system depends both on  $n_e$ and  on the electronic wave vector $q$ according to $\omega_p\propto \sqrt{n_e q}$ (see, for details, Sec.~\ref{SM:Plasma frequency in 2D systems}).
With increasing $n_e$, phonons will then be progressively screened starting with the short-period ones, thus yielding linewidth evolutions with $\Pexc$ compatible with Fig.~\ref{fig:FluenceDep}a and \ref{fig:FluenceDep}c. Furthermore, for $n_e\approx 10^{13}~$cm$^{-3}$, Fig.~\ref{SM:Plasma} shows that  $\omega_e$ exceeds the characteristic energy $k_B T=17.2$~meV for thermal phonons at 200~K even for the smallest in-plane wave vector $q=\pi/a_\mathrm{Trap}$ corresponding to the trap  ground state.
In fact, $\omega_p$ overcomes even the maximum LO phonon frequency for $q\geq\pi/5 a_\mathrm{Trap}$. 
In this context, high temperature polariton BEC becomes favored if the reduced phonon scattering overcompensates a potential reduction of the exciton oscillation strength by plasma screening. The latter is supported by the fact that  excitons are relatively insensitive to plasma screening~\cite{Fehrenbach_PRL49_1281_82}.
 
In conclusion, we have demonstrated that lateral confinement within $\mu$m-sized traps substantially increases polariton  stability in (Al,Ga)As structures and enables highly coherent polariton states -- polariton BEC condensates near room temperature. In addition to strong non-linearities, confined BECs also strongly interact with acoustic phonons and can enter self-oscillations  leading to phonon lasing~\cite{PVS333,PVS363}. These structures can thus be exploited for coherent optomechanics at GHz frequencies as well as for efficient sensing of solid-state excitations. Finally, the findings  pave the way towards applications of polariton-based functionalities at  room-temperatures using the (Al,Ga)As material system. Strategies towards this goal include  the use of larger lateral confinement barriers as well as of (Al,Ga)As QWs with higher excitonic stability.

\section{Methods}

\label{SECMethods}
\subsection {Structured (Al,Ga)As microcavities}

 The intracavity traps (cf. Fig.~\ref{FWHM_Rev}b and \ref{FWHM_Rev}c) were produced by combining  growth runs by molecular beam epitaxy (MBE) with patterning by optical lithography \cite{Kuznetsov_PRB97_195309_18} [see Sec.~\ref{subsec:SM_MC_Structure} for details].  The lower distributed Bragg reflector (DBR) and the MC spacer with GaAs QWs were grown in the first MBE run. The spacer embeds  two stacks of three 15-nm-thick GaAs QWs placed at antinodes of the  MC optical mode. The sample was then extracted from the MBE growth chamber and photolithographically patterned to defined shallow  (typically 10-20 nm-high)  $\mu$m-sized square mesas [cf.~Fig.~\ref{FWHM_Rev}b]. The upper DBR was then deposited in a second MBE run. The mesa shape remains imprinted in the surface profile during the MBE overgrowth and can be probed by atomic force microscopy~\cite{Kuznetsov_PRB97_195309_18}, as illustrated by  Fig.~\ref{FWHM_Rev}c. The variation of the spacer thickness results in a lower polariton energy in the non-etched $\mu$m-sized mesa regions resulting in traps with 
 with discrete polariton confined states~\cite{PVS312}. 
 Due to the temperature dependence of the material properties (i.e. GaAs band gap and refractive index), MC structures were redesigned and fabricated for each temperature range to ensure the appropriate energy matching between the excitonic and photonic states [see Sec.~\ref{subsec:SM_MC_Tdep} for details].

 
\subsection{Optical spectroscopy}
\label{Optical spectroscopy} 

The PL studies were performed under excitation by a single-mode, frequency stabilized laser beam (diameter of approx. $5~\mu$m) with energy tuned well above (approx. 100~meV) the polariton resonances. 
The PL was collected using a NA 0.26 microscope objective,  spectrally and spatially analyzed using a  grating spectrometer (resolution: 130~$\mu$eV) with a CCD detector. 
The nature of the polaritonic states and the Rabi splitting were determined by mapping the PL  along the surface of the MBE wafer (see, for details, Sec.~\ref{SM_HT_SCR}). 
The high-resolution studies were carried out by including a temperature-stabilized, piezo-tunnable Fabry-Perot etalon (FP) with a spectral resolution of 300~MHz in front of the spectrometer. 
 
\begin{acknowledgments}
We thank our technology team for the technical support in sample preparation as well as  Dr. P. John for a critical reading of the manuscript. We acknowledge the financial support by German DFG (grants 426728819 and 359162958).

\end{acknowledgments}


%




\clearpage
\beginsupplement
\widetext 

\begin{center}
    {\large \bf SUPPLEMENTARY INFORMATION:}\\ 
    \vspace{0.5 cm}
    {\large \bf Near-room-temperature zero-dimensional polariton lasers with sub-10 GHz linewidths}\\
    \vspace{0.5 cm}    
    {I.~dePedro-Embid$^{1*}$, A. S. Kuznetsov$^1$, K. Biermann$^{1}$, A. Cantarero$^{2}$,  
        and P. V. Santos$^{1\dagger}$}\\
    \it $^1${Paul-Drude-Institut für Festkörperelektronik, Leibniz-Institut im Forschungsverbund Berlin e. V., Hausvogteiplatz 5-7, 10117 Berlin, Germany}\\
    \it $^2${Molecular Science Institute, University of Valencia, PO Box 22085, 46071 Valencia, Spain}\\
    (corresponding authors: $^*$embid@pdi-berlin.de, $^\dagger$santos@pdi-berlin.de)
\end{center}


This supplement summarizes additional material  supporting and complementing the conclusions of the main manuscript.

\section{Material platforms for high-temperature polaritonics}
\label{sec:SM_Pol_FWHM_Rev}

In recent years, the pursuit of discovering novel solid-state phenomena and exploiting them for applications under ambient conditions has sparked significant interest in microcavity exciton polaritons able to operate in the strong coupling regime at room temperature. In this context, a wide range of novel semiconductor systems presenting robust exciton polaritons have emerged. Table~\ref{tab:Table1L} presents a comprehensive review of performance parameters  reported for polariton Bose-Einstein condensates realized in different material systems and operating at different temperatures. While room temperature polariton condensation has been observed across a broad range of material systems with large exciton binding energies, the coherence properties of the condensates (i.e.,  their spectral linewidths) far exceed those of the traditional GaAs-based systems at low temperature.  A selection of the data in Table~\ref{tab:Table1L} is plotted as a function of the BEC temperature in Fig.~\ref{FWHM_Rev}a of the main text. The table also includes information about the morphology and synthesis of the structures.

\begin{table*}[bhp]
\caption{\label{tab:Table1L} Performance parameters reported for polariton Bose-Einstein condensates reported for different material systems.
\footnote{Aclaration of the acronyms used in the table: 
$E_g$: band gap, $E_B$: exciton binding energy, $\Omega_R$: Rabi coupling,
MBE: Molecular beam Epitaxy, QW: Quantum Well, PLD: Pulsed Laser deposition, TDAF: 2,7-bis[9,9-
di(4-methylphenyl)-fluoren-2-yl]-9,9-di(4-methylphenyl)fluorene, CVD: Chemical Vapor Deposition, TMDCs: Transition Metal Dichalcogenides } }
\begin{ruledtabular}
\begin{tabular}{ccccccccc}
 \parbox{5em}{Material Type} & Material & Morphology & Synthesis & \parbox{3em}{$E_{g}$ at 300~K (eV)} & \parbox{3em}{$E_B$ (meV)} & \parbox{3em}{$2\Omega_{R}$ (meV)} & \parbox{3em}{BEC linewidth (meV)} & Ref.\\
 \hline
 \parbox{5em}{II-VI and III-V semiconductors}   
                                        &    GaAs    &    QW    &    MBE    &    1.422    &    10    &    5    & $2.5\times 10^{-3}$  (4K) - $\sim 1$ (70K) &    \citesupp{Deng_PNAS100_15318_03,Tempel_NJP14_83014_12}\\
                                        &             &         &           &             &          &    15 (10K) & 0.15 (10K)- NA (45K) & \citesupp{Bajoni_PRL100_47401_08}\\
                                        &    AlGaAs  &    QW    &    MBE    &    1.67     &    \parbox{5em}{> 26 (calculated)} & 14 (30k) - 8.5 (RT) & 1 (20K)  & \citesupp{Suchomel_OE25_24816_17}\\
                                        &    CdTe    &    QW    &    MBE    &    1.51     &    25    &    26   &   0.3 (5K)    &  \citesupp{Kasprzak_N443_409_06}\\
                                        &    GaN     &    Bulk  &    MBE    &    3.42     &    25    & 31 - 60 &         1 - 3 (RT)     &  \citesupp{Christopoulos_PRL98_126405_07}\\
                                        &            &     QW   &    MOVPE  &    3 - 3.3  &    40    & 17 - 56 &          0.5 (RT)      &  \citesupp{Christmann_APL93_51102_08}\\
                                        &            &           &          &    3.6      &    48    &    60   &        2-3 (4K-RT)     &  \citesupp{Delphan_AplPh8_021302_23}\\

                                        &    ZnO     & Microwire & Carbothermal &   3.31  &    60    & 200 - 300 &         > 2 (RT)     &  \citesupp{Zhang_PNAS112_E1516_15}\\
                                        &            &   Bulk    &     PLD      &   3.31  &    60    & 50 - 260  &          5 (RT)      &   \citesupp{Lu_OE20_5530_12}\\
                                        &    CdSe    & Nanoplatelets & Hot-injection method & 1.904 & 100-200 & 76 &     0.93 (RT)      & \citesupp{Yang_AS9_18_22}\\
 
\parbox{5em}{Organic semiconductors}    & Anthracene &   Single Crystal & \parbox{5em}{Melt-grown method} & 3.16 & 640  & 600-1000 & 1.5 (RT) & \cite{KenaCohen_NP4_371_10}\\  
                                        &    TDAF    &  Film     &   \parbox{5em}{Thermal evaporation} & 3.5 & NA & 600 - 1000 & 1 (RT) &   \citesupp{Daskalakis_NM13_271_14}\\

                                        &   \parbox{5em}{Conjugated polymer} & Film & Spin-coating & 2.7 & NA & 116 &      2 (RT)       &     \citesupp{Plumhof_NM13_247_13}\\
                                        & J-aggregate&   Film   &  Spin-coating & 1.8 & 360  & 50Fperovskites
                                        & 0.3 (200K) & \cite{Paschos_Sr7_11377_17}\\  

\parbox{5em}{Halide perovskites}        & $CsPbCl_{3}$ & Nanoplatelet & CVD     &  3.04  &    70     &  270      &      3.4 (RT)         &       \citesupp{Su_NL17_3982_17}\\
                                       
                                        & $CsPbBr_{3}$ & Nanoplatelet & CVD     &  2.407  &    40     &  120      &      1.3 (RT)       &   \citesupp{Su_SA4_18}\\
TMDCs                                   &  $WS_{2}$   & Flake         & exfoliation     &   1.9   &    710    &  37       &    2.4 (RT)        &    \citesupp{Zhao_NL21_3331_21, Zhu_SR5_9218_15}                               
\end{tabular}
\end{ruledtabular}
\end{table*}

\section{Polariton microcavities}

\subsection{Sample structure and fabrication} 
\label{subsec:SM_MC_Structure}

The studies were performed on  structured (Al,Ga)As polariton microcavities (MC) displayed in Figs.~\ref{FWHM_Rev}b and ~\ref{FWHM_Rev}c of the main text. The epitaxial (Al,Ga)As MCs were fabricated on a 2-inch (001) GaAs substrate via molecular beam epitaxy (MBE). In-situ, continuous spectral reflectivity spectra recorded during the epitaxial growth were used to achieve precise control of the growth parameters~\citesupp{Biermann_JCG557_125993_21}. 
 
 The sample structure consists of a spacer region sandwiched between a lower (LDBR) and an upper (UDBR) distributed Bragg reflector. Initially, a 14-period composition-graded LDBR was grown. Each period consists of a stack of three pairs of $ \mathrm{Al_{x_1}Ga_{1-x_1}As/Al_{x_2}As_{1-x_2}As}$ layers, each pair with an optical thickness of  $\lambda/4$ ($\lambda$ being the optical wavelength). The thicknesses $d_1$ and $d_2$ and Al compositions $x_1$ and $x_2$ of the layers pairs are listed in Table~\ref{tab:Table2} for a sample designed for  higher temperatures (sample $S_{HT}$, for $T\approx 225$~K) and Table~\ref{tab:Table3} for a low-temperature sample ($S_{LT}$, for $T\approx 77$~K).

Following the LDBR, the  MC spacer was deposited  including two stacks of three 15-nm-thick GaAs quantum wells (QWs) strategically placed at the antinode positions of the MC optical mode.
This special DBR structure also supports the confinement of low-frequency ($\sim 8$~GHz) acoustic phonons by utilizing three consecutive optical quarter-wavelenght layers as one acoustic quarter-wavelength layer~\cite{PVS354a}, a feature not exploited in this work. 
Subsequently, the sample was extracted from the MBE chamber to undergo patterning via photolithography and wet chemical etching. This etching process creates mesas of various shapes in the exposed spacer layer, inducing a lateral modulation of the cavity thickness and, consequently, of the cavity energy in the final structure.

The etching depth ( $\sim$ 17 nm) was chosen to blueshift the optical cavity mode in the etched areas (ER) by 9 meV (4.5 nm) as compared to the non-etched regions (nER). The upper surface of the etched layer was designed to correspond to a node of the optical cavity mode of the entire structure. This approach minimizes potential impacts of roughness or impurities introduced during ex-situ patterning on the optical properties of the structure. Furthermore, the shallow etching depth ensures that the QWs stay unaffected during the process, preserving their high quality. 

The sample was then reintroduced into the MBE system, subjected to cleaning through exposure to atomic hydrogen (30 minutes at 450 °C under a pressure of $2 \times 10^{-5}$ mbar), and subsequently overgrown with a $\lambda/4$ $\mathrm{Al_{0.10}Ga_{0.90}As}$ layer, followed by an 11-period composition graded UDBR with a layer structure that is the mirror image of the one for the LDBR.

The surface morphology of the overgrown sample was examined using atomic force microscopy (AFM) to check for possible imperfections or defects, as is shown in Fig.~\ref{FWHM_Rev}c of the main text.

\begin{table*}[bhp]
\caption{\label{tab:Table2} Detailed layer structure of the high temperature microcavity sample ($S_{HT}$, optimized for $T=225$~K). The DBRs consist of stacks of  $\mathrm{Al_{x_1}Ga_{1-x_1}As/Al_{x_2}As_{1-x_2}As}$ layers with thicknesses $d_1$ and $d_{2}$ and Al compositions $x_{1}$ and $x_{2}$, respectively. The total thickness ($d_T$) and the number of periods ($n_{rep}$) are also indicated.}
\begin{tabular}{|c|c|c|c|c|c|}
\hline
 {Region} & $d_T$ (nm) & $n_{rep}$ & $d_1  :  d_2$ (nm) & $x_1  :  x_2$ & Comment\\
 \hline
 UDBR     & 4268       &   11     & 64.88:59.76        &  0.50:0.10    &  pair$_{1}$\\
          &            &          & 69.46:59.76        & 0.90:0.10     &  pair$_{2}$\\
          &            &          & 69.46:64.88        & 0.90:0.50     &  pair$_{3}$\\
cavity etching& \textemdash & \textemdash & \textemdash & \textemdash  &  15 nm \\
Spacer    & 331        &   1      &                    &   0.10        &     2x3 15nm GaAs QWs\\

LDBR      &   5364     &   14     & 64.88:69.46         &   0.50:0.90  &  pair$_{3}$\\
          &            &          & 59.76:64.46         &  0.10:0.90   &  pair$_{2}$\\
          &            &          & 59.76:64.88         &  0.10:0.50   &  pair$_{1}$\\
Substrate &   350000   &          &                     &              & GaAs (001)\\                    
 \hline
 \end{tabular}

\end{table*}

\begin{table*}[bhp]
\caption{\label{tab:Table3}Detailed layer structure of the low temperature microcavity sample ($S_{LT}$, for $T=77$~K). The DBRs consist of stacks of  $ \mathrm{Al_{x_1}Ga_{1-x_1}As/Al_{x_2}As_{1-x_2}As}$ layers with thicknesses $d_1$ and $d_{2}$ and Al compositions $x_{1}$ and $x_{2}$, respectively. The total thickness ($d_T$) and the number of periods ($n_{rep}$) are also indicated.
}

\begin{tabular}{|c|c|c|c|c|c|}
\hline
 {Region} & $d_T$ (nm) & $n_{rep}$ & $d_1  :  d_2$ (nm) & $x_1  :  x_2$ & Comment\\
 \hline
 UDBR     & 4155       &   11     & 63.14:58.12        & 0.50:0.10    &   pair$_{1}$\\
          &            &          & 67.63:58.12        & 0.90:0.10     &  pair$_{2}$\\
          &            &          & 67.63:63.14        & 0.90:0.50     &  pair$_{3}$\\
cavity etching& \textemdash & \textemdash & \textemdash & \textemdash  &  15 nm \\
Spacer    & 289        &   1      &                    &   0.10        &     2x3 15nm GaAs QWs\\

LDBR      &   5289     &   14     & 63.14:67.63         &  0.50:0.90  &   pair$_{3}$\\
          &            &          & 58.12:67.63         &  0.10:0.90   &  pair$_{2}$\\
          &            &          & 58.12:63.14         &  0.10:0.50   &  pair$_{1}$\\
Substrate &   350000   &          &                     &              & GaAs (001)\\                    
 \hline
 \end{tabular}

\end{table*}

\subsection{Temperature dependence} \label{subsec:SM_MC_Tdep}

\begin{figure*}[tbhp]
    \includegraphics*[width=0.9\textwidth]{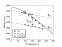}
    \caption
        {
           Temperature dependence of the bare heavy-hole exciton ($\mathrm{X_{hh}}$) and the bare cavity mode in the non-etched region ($\mathrm{C_{nER}}$) for the different studied samples: a high-temperature design ($\mathrm{S_{HT}}$, optimized for temperatures around 225~K), and a low-temperature design ($\mathrm{S_{LT}}$, optimized for temperatures around 77~K). The plot shows the strong temperature dependence of the detuning ($\delta_{CX}$). The points yield the energies obtained by fittings of the measured energies to a coupled oscillator model (cf. Fig.~\ref{fig:radialPL}. The solid line is the analytical calculation using the Varshni equation. The dashed lines correspond to linear fits for the temperature dependence of the bare cavity mode.  
        }
    \label{fig:SM_MC_Tdep}
\end{figure*}

In order to sustain polariton modes, the MC structure needs to be in the strong coupling regime. The latter requires that the energy of the bare cavity mode in the non-etched region ($\mathrm{C_{nER}}$) and the bare heavy hole energy of the QW ($\mathrm{X_{hh}}$) are kept close to each other (i.e., within the Rabi coupling energy). As the temperature rises, both $\mathrm{X_{hh}}$  and $\mathrm{C_{nER}}$ experience a red-shift due to the variation in GaAs band-gap and refractive index respectively. Notably, $\mathrm{X_{hh}}$ undergoes a more significant red-shift due to the small dependence of the refractive index with temperature compared with that of the band gap. Consequently, an increase in temperature causes the bare photonic and excitonic resonances to eventually drift apart until the strong coupling is lost. To address this issue, different MC structures must be designed for each target temperature.

For the studies carried out in this work, two different samples were grown targeting the strong coupling  at 77~K (sample $\mathrm{S_{LT}}$, cf.~Table~\ref{tab:Table3}) and 225~K (sample  $\mathrm{S_{HT}}$, cf.~Table~\ref{tab:Table2}). The QW stacks  were kept the same for both samples while the DBR structure was modified. Figure~\ref{fig:SM_MC_Tdep} shows the measured energies of the heavy-hole QW exciton and the bare cavity mode in the non-etched region as a function of temperature. The points correspond to experimentally measured values, obtained by fitting to a coupled oscillator model, as will be explained in detail in section \ref{SM_HT_SCR}. The solid line is an analytical calculation assuming a temperature-dependence of the band-gap following the Varshni equation~\citesupp{VARSHNI1967149}. The dashed lines correspond to linear fits for the temperature dependence of the bare cavity mode. 

It is important to emphasize that the  dependence on temperature of the  photon-to-exciton detuning ($\mathrm{\delta_{CX}}$) is markedly different in the high temperature range studied here (150-200~K) compared to the typically studied low-temperature regime (4-70~K). For temperatures below 70~K the temperature dependence of GaAs band gap is small and  $\mathrm{\delta_{CX}}$ remains almost constant. This is in contrast to the behavior at  temperatures above 150~K (see Fig~\ref{fig:SM_MC_Tdep}), where the temperature dependence of $\mathrm{\delta_{CX}}$ becomes very pronounced. 

\section{Strong coupling up to 225K}
\label{SM_HT_SCR}

\begin{figure*}[tbhp]
\includegraphics*[width=\textwidth]{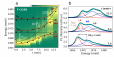}
\caption
    {
         (a) Radial ($r$) dependence of the polariton PL in a planar MC recorded at 225~K. The black circles mark the PL peak energies obtained from a multi-peak Lorentzian fit to each spectrum  as a function of $r$. The solid lines are fits to a photon-exciton coupled-oscillator model for the interaction between the bare optical MC mode (C) and the QW heavy-hole ($X_{hh}$) and light-hole ($X_{lh}$) exciton. A clear anti-crossing behavior is observed characterized by a Rabi splitting energy of $2\Omega_R = 4$ meV (equal to the minimum energy separation between the middle (MP) and lower polariton (LP) branches). The dashed lines correspond to the bare photon and exciton energies, showing that the detuning ($\delta C_{X}= X_{hh}-C$) can be changed by selecting  different radial positions along the sample.  
         (b) Spectral PL profiles corresponding to the vertical cross-section at different radii $r$  of the map in (a). The solid lines are Lorentzian fits to the emission of the LP (red), MP (blue), and UP (black) displaying the anti-crossing behavior around $r=7.6$~mm. 
    }
\label{fig:radialPL}
\end{figure*}

{
The polaritonic nature of the  states confined in the traps and their boundary states as well as the light-matter coupling energy was determined by mapping the polariton energies along the surface of the MBE wafer.  
Figure~\ref{fig:radialPL}a displays the radial ($r$) dependence of the photoluminescence (PL) recorded on a 2'' MBE wafer with a MC designed for 225~K. The  black dots mark  the center energies of the three resonances identified in the plots corresponding to the lower (LP), middle (MP), and upper polariton (UP) levels (some of these resonances cannot be identified in the color scale of the figure).  
In this structure, the photonic cavity mode (C) at $r = 0$ is red-shifted by 4~meV with respect to the bare electron-heavy hole exciton in the QWs. Due to non-uniformities of the MBE fluxes on the wafer surface, the thicknesses of the MBE layers decrease by approx. 1.5~\% as one moves from the center ($r=0$) towards the edges of the 2'' wafer. This reduction blue-shifts the bare optical mode of the MC (C), while the bare electron-heavy hole ($X_{hh}$) and electron-light hole ($X_{lh}$) resonances remain approximately constant at  $E_{hh}=1470$ and $E_{lh}=1476$~meV, respectively. Moving along the radial direction thus scans the bare photon energy across the excitonic resonances leading to the anti-crossing behaviors at $r=7.6$~mm and $r=11$~mm characteristic of polariton formation. 
 The solid lines in Fig.~\ref{fig:radialPL}(a) are fits to a photon-exciton coupled oscillator model for the coupling between the bare optical MC mode and the heavy-hole $X_{hh}$ and light-hole $X_{lh}$ excitons taking into account the variations in layer thicknesses, which yield the bare resonance energies (dashed lines) as well as the effective Rabi splitting energies of $2\Omega'^{hh}_R = 4$ meV and $2\Omega'^{lh}_R = 3$ meV for the $X_{hh}-C$ and $X_{lh}-C$ couplings, respectively. 
}

    Figure~\ref{fig:radialPL}(b) shows spectral PL profiles corresponding to vertical cross-sections along different radii $r$  of the map in  Fig.~\ref{fig:radialPL}(a). The solid lines are Lorentzian fits to the emission of the LP (red), polariton (blue), and UP (black) displaying the anti-crossing behaviors around $r=7.6$~mm and $r=11.3$~mm. The latter gives additional evidence for the strong exciton-photon coupling leading to polariton formation at these temperatures. 
    The strongly photonic LP level at $r=0$ in Fig.~\ref{fig:radialPL}(a) blue-shifts and broadens as its excitonic content increases for increasing $r$. The effective Rabi splitting  $2\Omega'^{hh}_R=4$~meV measured above for 225~K is significantly lower than the one  measured in similar structures designed for 10~K ($2\Omega'^{hh}_R = 8$ meV). This reduction is attributed to the larger excitonic linewidths $\gamma_X$ at 225~K, which makes the effective Rabi splitting much smaller than twice the real Rabi coupling $2\Omega^{hh}_R$ according to:~\citesupp{Savona_SSC93_733_95}

   \begin{equation}
   \Omega'^{hh}_R\approx \sqrt{(\Omega^{hh}_R)^2-\frac{1}{2} ( \gamma_X^2 + \gamma_C^2)}\end{equation}
   
   \noindent By taking $\gamma_X \sim 4$~meV calculated from $\gamma_X =  a_{ph}T + b_{ph}N_{LO}(T)$ \citesupp{Gopal_JAP87_4_00}  and $\gamma_C \sim 0.3$~meV from reflectivity measurements at low temperature (and assuming that $\mathrm{\gamma_C}$ doesn't depend on temperature),   we estimate $\Omega'^{hh}_R\approx 7$~eV. In contrast, for 10~K $\gamma_X\ll\Omega^{hh}_R$ and $\Omega'^{hh}_R=\Omega^{hh}_R \approx 8$ ~meV. The light-matter coupling energy thus remains approximately the same for the two temperatures.

\section{Dependence of the emission energy on excitation power} 
\label{sec:SM_Pexc}


\begin{figure*}[tbhp]
    \includegraphics*[width=\textwidth]{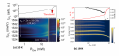}
    \caption
        { Optical excitation power ($\Pexc$) dependence of confined polaritons at (a) 10~K and (b) 190~K. The upper panels show the integrated PL intensity.}
    \label{fig:excDepTemp}
\end{figure*}
    
Figures~\ref{fig:excDepTemp}(a) and \ref{fig:excDepTemp}(b) compare the dependence of the emission lines for  (a) a $4\times 4~\mu$m$^2$ trap at 10~K and (b) a $5\times 5~\mu$m$^2$ trap at 190~K on the excitation fluence $\Pexc$.  
The experiments were performed on traps with different detunings  $\delta_{CX}$'s. We note that the threshold excitation fluences $\Pth$ for condensation increases from approx. 30 mW at 10~K  to over 60~mW at 190~K.  
For both temperatures, the emission energies of the different confined states at low excitation densities blue-shifts with increasing fluences: this behavior is attributed to increased repulsive polariton-polariton interactions with increasing particle densities. At low temperatures, the trap emission initial blue-shifts with increased excitation: beyond the condensation threshold ($\Pth$), the emission energy remains, however,  approximately constant with excitation density [a red-shift in the condensation regime can many times be observed, but for much higher excitation powers, see, e.g., Fig.~2a of Ref.~\onlinecite{Chafatinos_NC14_3485_23}]. 
For the high temperature sample of Fig.~\ref{fig:excDepTemp}(b), in contrast, the blue-shift is much smaller than at low temperatures, despite the higher excitation densities. In addition, it turns into a significant red-shift, which  start well  below the threshold fluence and continues over the condensation regime.

\section{Electronic nature of the red-shifts at high excitation}
\label{sec:SMTEffects}

\begin{figure}[tbhp]
    \centerline{\includegraphics[width=0.5\textwidth]{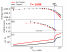}}
    \caption
        {
            Dependence of the (a) energy, (b) linewidth (FWHM) and (c) PL intensity of the confined ground 
            state of  Fig.~\ref{fig:excDepTemp}(b) on the excitation power $\Pexc$ recorded for two different experimental conditions: continuous (i.e., unchopped) excitation(red) and chopped excitation with a duty cycle of ~1~\% and a repetition rate of 65~Hz (black). The vertical blue line marks the threshold power ($\Pth$)  
        }
    \label{fig:PLdutycycle}
\end{figure}

One might argue that the sub-threshold red-shifts originate from the thermal loading under the higher excitation fluences 
required for lasing [cf.~Figs.~\ref{fig:BECCoherence}b  and \ref{fig:excDepTemp}(b) of the main text].
In particular, heating effects become accentuated by the  strong reduction of the thermal conductivity of GaAs (by a  factor of approx. 40 times between 10~K and 225~K,\citesupp{Carlson_JAP36_2_65}). 
%
To refute this possibility,  we have conducted additional PL experiments with a chopper inserted in the laser path to reduce the illumination duty cycle. This setup allowed us to compare the PL response under continuous optical excitation (i.e., with a 100\% duty cycle) with the one, where the laser excitation pulses are applied with a small duty cycle (of approximately 1\% at a repetition rate of 65~Hz). Figure\ref{fig:PLdutycycle} compares the excitation fluence dependence of the energy, linewidth and PL intensity of the first confined state recorded under these two conditions. We find that the observed trends in energy and linewidth are only marginally affected by the low-duty cycle. Apart from the expected intensity offset of the integrated PL intensity (which is a direct consequence of the different densities excited at the different duty cycles), a similar behavior is observed for both cases. Notably, the characteristic super-linear increase of the PL intensity marking the onset of polariton condensation appears at the same threshold excitation fluences ($Pth$), as depicted by the blue vertical line in  Fig.~\ref{fig:PLdutycycle}(c).

These results prove that the red-shift observed at high temperatures and fluences is mainly caused by the high density of particles created by the intense laser beam rather than by the thermal heating induced by laser absorption. We thus attribute the redshift to variations of the refractive index induced by the high particle density.

\section{Photoluminescence instabilities in the lasing regime}
\label{sec:SMInstabilities}

\begin{figure*}[tbhp]
    \includegraphics*[width=0.95\textwidth]{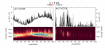}
    \caption
        {
            Photoluminescence instabilities at high temperatures of a  $5\times 5~\mu$m$^2$ trap  at 
            a. 200~K and 
            b. 225~K recorded as successive acquisition frames by a CCD camera with an acquisition time of 100~ms per frame. The bottom panels show a color map of the PL emission energy as a function of time with the PL intensity encoded in colors. The top panels display the integrated PL intensity. 
        }
    \label{fig:PLFluctuations}
\end{figure*}

The emission of the polariton condensates at low temperatures (up to $T\approx 100$~K) is remarkably stable with linewidths below 2~GHz (cf. Fig.~\ref{fig:BECCoherence} of the main text, see also Ref.~\onlinecite{PVS354a}). Above 150~K, the experiments proved to be challenging due to the required high-level of temperature control as well as for the onset of instabilities in the emission intensity and energy for excitations densities close or above  $P_\mathrm{th}$.
These instabilities  could be  minimized (but, in most cases, not completely eliminated) by excitation using a single mode, frequency stabilized pump laser. 


Figure~\ref{fig:PLFluctuations} provides a comprehensive summary of the phenomenology of these fluctuations. Notably, the fluctuations depend on the excitation conditions and impact not only the emission intensity of the condensate [top panels of Fig.~\ref{fig:PLFluctuations}a-b] but also the emission energy (bottom panels of Fig.~\ref{fig:PLFluctuations}a and \ref{fig:PLFluctuations}b). The coupling between intensity and energy fluctuations is attributed to the highly nonlinear nature of the polariton system, where small changes in the particle density directly affect the emission energy, which then feedbacks by modifying the particle injection efficiency. The energy fluctuations also contribute to the peculiar lineshape of the high-resolution PL spectra of Fig.~\ref{fig:BECCoherence}b of the main text.

Interestingly, the amplitude and shape of the PL instabilities is temperature-dependent. At temperatures up to 200~K, as depicted in Fig.~\ref{fig:PLFluctuations}a, the fluctuations normally appear on top of a continuous PL background. For higher temperatures (Fig.~\ref{fig:PLFluctuations}b), in contrast, the amplitude of the instabilities increases while the one of the background background reduces.
%
%
%
Further measurements (not shown)  reveal the presence of additional frequency components oscillating on a millisecond timescale, indicating the complexity of the oscillation process. The interplay between the low-frequency and high-frequency oscillations likely points to a mechanism combining  optoelectronic and thermal effects.

Optomechanical instabilities and self-oscillations with frequencies in the hundred of MHz range  have been recently reported for GaAs disk resonators~\citesupp{Allain_PRL_126_243901_21} and related to the mutual coupling between light, mechanical oscillations, carriers, and heat exchange. In contrast, the instabilities observed here are random with much longer time scales (typically exceeding  a few ms).
Although the condensates were non-resonantly excited well above (i.e., $>100$~meV) the PL energies,  
their emission is very sensitive to small changes in the excitation wavelength and power. The associated fluctuations in particle density may, by themselves, induce small variations in temperature and/or the particle injection efficiency, favoring an unstable behavior. The role of thermal effects can be estimated from the temperature dependence of the GaAs band gap at 200~K (of approx. 500~$\mu$eV/K), which is much stronger (by a factor of 40)  than at cryogenic temperatures. From this dependence,  one can easily estimate that a temperature shift of only $\sim 0.2$~K yields already a polariton energy shift of 10~GHz, which exceeds the linewidth of the highly coherent BEC state of Fig.~\ref{fig:BECCoherence} of the main text.
A comprehensive understanding of these mechanisms requires further investigations. 

\label{sec:SM_eff_mass_cont}
\begin{figure*}[tbhp]
    \includegraphics*[width=0.7\textwidth]{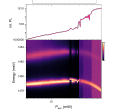}
    \caption
        {  Excitation power ($\Pexc$) dependence of the emission at 180~K of a $5\times 5~\mu$m$^2$ trap with a detuning of a heavy-hole exciton Hopfield coefficient  $0.07$. 
%
        The upper panel displays the integrated PL emission from the GS. }
    \label{fig:SM_eff_mass_cont}
\end{figure*}

\section{Evolution of the polariton energy accross the lasing threshold}
\label{sec:Continuity of the polariton energy accross the lasing threshold}

A direct way to assess the polaritonic or photonic nature of the polaritons across the lasing transition consists in measuring the energy dispersion (or, equivalently, the effective mass of the particles). In the case of confined polaritons, the momentum $\hbar k$ of the states is discrete, so that it is not possible to get the full wave vector ($k$) dependence. The energy of the confined states, however, depends on their effective mass. If the system transitions from strong coupling to the weak coupling regime at the lasing threshold, the effective mass should reduce and, as a result, the energy of the confined levels would shift, giving rise to a discontinuity in energy. This effect is expected to be  more pronounced for polaritons with a high excitonic fraction, which have masses much larger than the effective photonic mass in the MC.
We show in Fig.~\ref{fig:SM_eff_mass_cont}  the excitation power dependence of a trap with a high   excitonic content exhibiting a transition to the lasing regime. The Hopffield coefficients ($\mathrm{H_X^2}$) of the ground state (GS) and first excited state (ES) are estimated to be 0.07 and 0.19 respectively. The effective mass $M_{LP}^*$ of the polaritons levels can be calculated as 

\begin{equation}
    \frac{1}{M_{LP}^*} \approx \frac{1-H_X^2}{M_C^*},
\end{equation}

\noindent where we have considered $H_C^2/M_X^* \approx 0$ since $M_X^* \gg M_C^*$. The effective mass of the cavity can be also estimated using the expression
 \begin{equation}
     M_C^*=m \frac{\pi n_c \hbar}{c_o L_c}.
 \end{equation}
 
 \noindent In this expression, $c_0$ is the light speed in vacuum and $\hbar$ Planck's constant. Substituting  the sample parameters for the cavity mode of order ($m=3/2$), the cavity length ($L_c =350$~nm) and refractive index ($n_c \approx 3.6$),  we estimate an  effective mass for the cavity mode of $M_C^*\approx 2 \cdot 10^{-5} m_e$.

 We can now estimate the change in the energy of the confined levels when the effective mass goes from $M_{LP}^*$ to $M_C^*$, i.e., if the system transitions from the strong to the weak coupling regime. For the GS we get:
 \begin{equation}
     \Delta E_{GS}= \frac{\hbar^2 \pi^2}{m_e a_\mathrm{Trap}^2} \left( \frac{1}{M_{LP}^*}-\frac{1}{M_C^*} \right) [\text{meV} \cdot 10^{-5}m_e]
 \end{equation}

 By substituting the values for the effective masses and the trap side length ($a_\mathrm{Trap}=2~\mu$m), we calculate the energy disconuity at the threshold to be $\Delta E_{GS} \approx 0.6$~meV for the GS. A similar calculations predicts an energy shift $\Delta_{ES}$ for the ES given by:
 
 \begin{equation}
     \Delta_{ES}=\frac{5}{2} \frac{\hbar^2 \pi^2}{m_e a_\mathrm{Trap}^2} \left( \frac{1}{M_{LP}^*}-\frac{1}{M_C^*} \right) [\text{meV} \cdot 10^{-5}m_e],
 \end{equation}
 
 \noindent yielding $\Delta_{ES} \approx 4.7$~meV. 
It is clear from the data displayed in Fig.~\ref{fig:SM_eff_mass_cont} that such a big discontinuities are not experimentally observed in our system, thus giving evidence that the effective mass more or less stays constant accross the threshold transition, i.e., that the strong coupling is maintained across the transition with no transition to the weak coupling regime.

\begin{figure}[tbhp]
     \centerline{\includegraphics[width=0.7\textwidth]{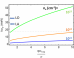}}
    \caption{ Two-dimensional plasmon dispersion  $\hbar \omega_p \times q$ in a GaAs QW as determined from Eq.~\ref{SM:EqPlasma2} for the indicated free electron densities, $n_e$. 
    The wave vectors $q$ are normalized to $\pi/a_\mathrm{Trap}$, where $a_\mathrm{Trap}$ is the lateral trap size.
    The vertical dashed line marks the characteristic thermal energy $k_B T=17.2$~meV for thermal phonons at 200K. The arrows mark the maximum energy of LO and LA phonons of GaAs. The calculations were carried out  using the following parameters:  $c_0=3\times 10^8$~m/s, 
     $ e=1.602\times 10^{-19}$~C,  $\varepsilon_b=16\varepsilon_0$ with  $\varepsilon_0=8.85\times 10^{-12}$~F/m, and $m_e =  m_e^* m^0_\mathrm{e}$, where  $m_e^*=0.067$ is the electron effective mass in GaAs and $m^0_\mathrm{e}= 9.11\times 10^{-31}$~kg the free electron mass. 
     }
     \label{SM:Plasma}
    \end{figure}
    
\section{Plasma frequency in 2D systems}
\label{SM:Plasma frequency in 2D systems}

The interaction with the long-range electrical field of polar acoustic and optical phonons is a dominant phonon scattering mechanism in GaAs structures. The ability of free carriers to screen this piezo electrical phonon field is mainly dictated by the plasma frequency $\omega_p$ of the free carriers, which increases with increasing carrier density $n_e$. The piezoelectric  field of phonons with frequencies below the plasma frequency is effectively screened by a dynamic rearrangement of the carrier populations. In contrast, the carrier density cannot readjust itself to respond to phonons with frequency above $\omega_p$, which prevents an effective phonon screening. 

The trapping polariton potential confines not only photons but also phonons: the relevant long wavelength polar phonons acting on on the polariton excitonic component are thus expected to have lateral wave vector components $q_{ph}\sim m \pi/a_\mathrm{Trap}$, where $m=1,2,\dots$ and $L_\mathrm{Trap}$ is the lateral trap size. While in 3D systems  $\omega_p$ only depends on the carrier density, in 2D it is a function of both the areal density, $n_e$, and the carrier wave vector $q$. We will consider in the following the case of electrons, which, due to the smaller effective mass, have plasma frequencies higher than for holes. 

The relationship between $\omega_p$ for electrons with a density $n_e$ in a 2D system (such as, e.g., a QW)  and the wave vector $q$ can be expressed in SI units by \citesupp{Stern_PRL18_546_67}:

\begin{equation}
  q^2=\frac{\varepsilon_b\omega_p^2}{\varepsilon_0c_0^2}+\left(\frac{\omega^2_p}{a_p}\right)^2,
  \label{SM:EqPlasma1}
\end{equation}

\noindent where $\varepsilon_b$ is the background (static) dielectric constant (i. e., $\varepsilon_b =\varepsilon_r \varepsilon_0$, where $\varepsilon_0$ is the vacuum permitivity and $\epsilon_r$ is the relative permitivity of the medium).
The parameter $a_p$ is related to the electron charge ($e$) and  mass ($m_e$)  according to: 

\begin{equation}
  a_p=\frac{ n_e e^2}{ 2 m_e\varepsilon_b}.
\label{SM:Eqap}
\end{equation}

\noindent The second order Eq.~\ref{SM:EqPlasma1} can be solved for $\omega_p$ to give:

\begin{equation}
  \omega_p=\sqrt{\frac{\varepsilon_r a_p^2}{2c_0^2}\left(\sqrt{1+\frac{4c_0^4q^2}{a_p^2\varepsilon_r^2}}-1\right)}
\label{SM:EqPlasma2}
\end{equation}

The last equation can be simplified to yield approximations for $\omega_p$ in two asymtopic limits. If $q$ is large, one obtains

\begin{equation}\label{SM:largeq}
  \omega_p\approx\sqrt{a_pq},
\end{equation}

\noindent thus yielding the acoustic  plasma dispersion with the plasma frequency increasing with $\sqrt{q}$. In the opposite limit (i. e., for small $q$), one obtains a linear dispersion given by:

\begin{equation}
\label{SM:qsmall}
  \omega_p\approx \frac{c_0q}{\sqrt{\varepsilon_r}}.
\end{equation}

\noindent The latter  corresponds to the dispersion of electromagnetic waves propagating in a medium with dielectric constant $\varepsilon_b$.

Figure~\ref{SM:Plasma} displays the plasmon dispersion calculated using Eq.~\ref{SM:EqPlasma2} for  a GaAs QW with different free electron densities $n_e$. 
The wave vectors $q$ are normalized to $\pi/a_\mathrm{Trap}$. For a laterally confined state,  $q$ is quantized in units of $\pi/a_\mathrm{Trap}$, where $a_\mathrm{Trap}$ is the lateral trap size. 
For small densities, the dispersion in Fig.~\ref{SM:Plasma} is approximately linear following Eq.~\ref{SM:qsmall}. For high densities $n_e =10^{13}$~cm$^{-2}$, the plasmon energy follows the square-root dependence given by Eq.~\ref{SM:largeq} for the whole range of $q$'s displayed in the figure. Furthermore, the plasma energy at $n_e =10^{13}$~cm$^{-2}$ exceeds the characteristic thermal energy $k_B T=17.2$~meV for phonons at 200~K (dashed black line) even for the smallest electronic wave vector ($qa_\mathrm{Trap}/\pi=1$). 
 For relatively small wave vectors $qa_\mathrm{Trap}/\pi\geq5$, the plasma frequency also exceeds the maximum phonon energy in the host material (indicated by the arrows in the figure).
  We conclude, therefore, that a high-density of photo-excited electron-hole pairs can efficiently screen polar phonons.







\end{document}